\documentclass[preprint,prc,aps,showpacs,groupedaddress,floatfix]{revtex4}
\usepackage{dcolumn}
\usepackage{bm}
\usepackage{graphics}
\usepackage{graphicx}
\usepackage{epsfig}
\usepackage{amssymb}
\usepackage{amsmath}
\begin{document}

\title{An Improvement of the Asymptotic Iteration Method for Exactly Solvable Eigenvalue Problems\footnote{Supported by the Scientific
and Technical Research Council of Turkey (T\"{U}B\.{I}TAK) under
Grant No TBAG-106T024, Erciyes University (E\"{U}BAP-FBA-04-02), and
Turkish Academy of Sciences (T\"{U}BA-GEB\.{I}P).}}
\author{I. Boztosun\footnote{Email: boztosun@erciyes.edu.tr}, M. Karakoc}
\affiliation{Faculty of Arts and Sciences, Department of Physics,
Erciyes University, Kayseri, Turkey}

\date{(Received 27 July 2007)}

\begin{abstract}
We derive a formula that simplifies the original
asymptotic iteration method formulation to find the energy
eigenvalues for the analytically solvable cases. We then show that
there is a connection between the asymptotic iteration and the
Nikiforov--Uvarov methods, which both solve the second order linear
ordinary differential equations analytically.
\end{abstract}
 \maketitle

Analytical solutions of the radial Schr\"{o}dinger equation are of
high importance in quantum mechanics since the wave function
contains all the necessary information to describe a quantum system
fully$^{[1-12]}$. The asymptotic iteration method (AIM)$^{[1,2]}$
and the Nikiforov--Uvarov method$^{[3]}$ are two different
approaches to solve the resulting Schr\"{o}dinger equation and they
have been successfully applied to a wide variety of
problems$^{[4-14]}$.

In this work, we derive a formula which simplifies the original AIM
formulation to find the energy eigenvalues for the analytically
solvable cases. This formulation shows that there is a connection
between the AIM and Nikiforov--Uvarov methods regarding the solution of
differential equations.

AIM$^{[1]}$ is proposed to solve the second-order differential
equations in the form
\begin{equation}\label{diff}
  y''=\lambda_{0}(r)y'+s_{0}(r)y,
\end{equation}
where $\lambda_{0}(r)\neq 0$ and s$_{0}$(r), $\lambda_{0}$(r) are in
C$_{\infty}$(a,b). Eq. (1) as given in Ref.\,[1] has a general
solution
\begin{equation}\label{generalsolution}
  y(r)=exp \left( - \int^{r} \alpha(r^{'}) dr^{'}\right ) \left [C_{2}+C_{1}
  \int^{r}exp  \left( \int^{r^{'}} (\lambda_{0}(r^{''})+2\alpha(r^{''})) dr^{''} \right ) dr^{'} \right
  ],
\end{equation}
if $k>0$, for sufficiently large $k$, $\alpha(r)$
values can be obtained by
\begin{equation}\label{alphak}
\frac{s_{k}(r)}{\lambda_{k}(r)}=\frac{s_{k-1}(r)}{\lambda_{k-1}(r)}=\alpha(r),
\end{equation}
where
\begin{eqnarray}\label{iter}
  \lambda_{k}(r) & = &
  \lambda_{k-1}'(r)+s_{k-1}(r)+\lambda_{0}(r)\lambda_{k-1}(r), \quad
  \nonumber \\
s_{k}(r) & = & s_{k-1}'(r)+s_{0}(r)\lambda_{k-1}(r), \quad \quad
\quad \quad k=1,2,3,\cdots
\end{eqnarray}

In this method, the idea is to convert the radial Schr\"{o}dinger
equation into the form of Eq.\,(1) for a given potential. Then,
s$_{0}(r)$ and $\lambda_{0}(r)$ are determined, and s$_{k}(r)$ and
$\lambda_{k}(r)$ parameters are calculated. The energy eigenvalues
are determined by the quantization condition given by the equation:

\begin{equation}\label{quantization}
  \delta_{k}(r)=\lambda_{k}(r)s_{k-1}(r)-\lambda_{k-1}(r)s_{k}(r)=0, \quad \quad
k=1,2,3,\cdots.
\end{equation}

If we rearrange Eq.\,(1) by introducing new forms for $\lambda_0(r)$
and $s_0(r)$ as follows:
\begin{equation}\label{aimnu1}
  \lambda_0(r)=-\frac{\tau(r)}{\sigma(r)},\,\quad s_0(r)=-\frac{\gamma_n}{\sigma
  (r)},
\end{equation}
where $\tau$ and $\sigma$ are new functions and $\gamma_n$ is a
constant with which we quantize the energy. Inserting them into Eq.\,(1),
the second-order differential equation  takes the form
\begin{equation}\label{aimnu2}
  y''=-\frac{\tau(r)}{\sigma(r)}y'-\frac{\gamma_n}{\sigma
  (r)}y .
\end{equation}
An essential ingredient of AIM is the quantization condition given
by Eq. (5). Every particular value in Eq. (5) corresponds to an
energy eigenvalue. Therefore, if we first take $k=1$ and $n=0$, put
them into Eq. (6) and then use Eq. (4) for $\gamma_0 \neq 0$, the
quantization condition given by Eq. (5) reduces to
\begin{equation}\label{aimnu3}
  \tau^{'}(r)+\gamma_0=0,
\end{equation}
which should hold for all $r$. This means that $\tau(r)$ is a
polynomial of the first degree at most. If we choose now $n=1$ and
$k=2$, using Eq. (4), the quantization condition given by Eq. (5)
reduces to
\begin{equation}\label{aimnu4}
\left[\gamma_1+\tau^{'}(r) \right ]\left[(\gamma_1+2\tau^{'}(r)+\sigma^{''}(r))-\tau^{''}(r)(\tau(r)+\sigma^{'}(r)) \right ]=0,
\end{equation}
which should also hold for all $r$. Since we have already
established that $\tau(r)$ is a polynomial of the first degree at
most, thus, $\tau{''}(r)=0$. Assuming the spectrum of our problem to
be non-degenerate, \emph{i.e.} $\gamma_1 \neq \gamma_0$, and taking
into account Eq. (8), we conclude that $\sigma{''}(r)$ is a
polynomial of the second degree at most. In such a case, Eq. (7)
and, hence, Eq. (1) are of hypergeometric type (We point out that
the comments in this part have been made by an unanimous referee of
the Phys. Lett. A for another paper of ours).

Already knowing the properties of $\tau(r)$ and $\sigma(r)$ and by
using AIM procedure given by Eq. (4), we can calculate
$\lambda_n(r)$ and $s_n(r)$ as follows, in order to formulate the
energy eigenvalue equation,
\begin{eqnarray}\label{aimnu5}
  &&\lambda_1(r)=\frac{\gamma \left(\tau (r)+\sigma
   '(r)\right)}{\sigma (r)^2},\,\quad
  s_1(r)=\frac{\tau (r) \left(\tau (r)+\sigma '(r)\right)-\sigma
   (r) \left(\gamma+\tau '(r)\right)}{\sigma (r)^2},\\
   &&\cdots \emph{etc} \nonumber
\end{eqnarray}
If we use the termination condition of AIM given by Eq. (5), the
energy eigenvalues are obtained as follows
\begin{eqnarray}
\frac{s_0 }{\lambda _0 } &=& \frac{s_1 }{\lambda_1}\,\Rightarrow
\,\ \gamma_0  =  0 \\
 \frac{s_1 }{\lambda _1 } &=& \frac{s_2 }{\lambda _2 }\,\Rightarrow
\,\ \gamma_1 =  -\tau '(r) \\
 \frac{s_2 }{\lambda _2 } &=& \frac{s_3 }{\lambda _3 }\,\Rightarrow
\,\ \gamma_2 =  -2 \tau '(r)-\sigma ''(r) \\
 \frac{s_3 }{\lambda _3 } &=& \frac{s_4 }{\lambda _4 }\,\Rightarrow
\,\ \gamma_3 =  -3 \tau '(r) - 3 \sigma ''(r), \\
\cdots \emph{etc} \nonumber
\end{eqnarray}
which can be generalized as
\begin{equation} \label{eq19}
\gamma_n = - n\tau '(r) - \frac{n(n-1)}{2}\sigma ''(r)
\end{equation}
This formula gives directly the energy eigenvalues for
differential equation (7). It replaces the
iterative calculations needed to find the energy eigenvalues in the
original AIM formulation. Also, it is exactly the same as the Nikiforov--Uvarov energy
eigenvalue equation. This is briefly derived as follows:

The Nikiforov--Uvarov equation is written as
\begin{equation} \label{nu1}
\psi (r{)}'' + \frac{{\tau }(r)}{\sigma
(r)}{\psi }'(r) + \frac{\tilde {\sigma }(r)}{\sigma ^2(r)}\psi (r)
= 0,
\end{equation}
where $\sigma (r)$ and $\tilde {\sigma }(r)$ are the polynomials of
the second-degree at most, and $\tilde {\tau }(r)$ is a first-degree
polynomial$^{[3,4,6-9]}$. Hence, from Eq. (16), the Schr\"{o}dinger
equation or the Schr\"{o}dinger-like equations can be solved
analytically by this method. In order to find a particular solution
of Eq. (16), the following transformation is used:
\begin{equation}
\psi (r) = \phi (r)y(r).
\end{equation}
It reduces Eq. (16) to an equation of hypergeometric type,
\begin{equation}
\label{nu2} \sigma (r){y}'' + \tau (r){y}' + \gamma y = 0,
\end{equation}
and $\phi (r)$ is defined as a logarithmic derivative in the
following form and its solutions can be obtained from
\begin{equation}
\label{nu3} {{\phi }'(r)} \mathord{\left/ {\vphantom {{{\phi
}'(r)} \phi }} \right. \kern-\nulldelimiterspace} \phi (r) = {\pi
(r)} \mathord{\left/ {\vphantom {{\pi (r)} {\sigma (r)}}} \right.
\kern-\nulldelimiterspace} {\sigma (r)}
\end{equation}
The function $\pi $ and the parameter $\lambdabar $ required for
this method are defined as follows:
\begin{equation}
\label{nu4} \pi (r) = \frac{{\sigma }' - {\tau }}{2}\pm
\sqrt {\left( {\frac{{\sigma }' - {\tau }}{2}} \right)^2 -
\tilde {\sigma } + k\sigma },
\end{equation}
\begin{equation}
\label{nu5} {\lambdabar} = k + {\pi }'.
\end{equation}
On the other hand, in order to find the value of $k$, the
expression under the square root must be the square of a
polynomial. Thus, the eigenvalue equation for the
Schr\"{o}dinger equation becomes
\begin{equation}
\label{nueigen} \lambdabar_n = - n{\tau }' - \frac{n(n -
1)}{2}{\sigma }''.
\end{equation}

Next, we present some applications. For classical differential
equations, in order to show the applicability of the above formula,
let us consider the classical differential equations which give
polynomial solutions$^{[15]}$. The results are presented in
Table\,1.

For some other applications reported previously in the literature
by using AIM, we may take the following examples:

In the Morse potential case, if we take the transformed equation
(31) of Ref.\,[16] as
\begin{equation}\label{aimschrm}
\frac{d^{2}f_{n}(r)}{dr^{2}}=\left(\frac{2\beta y-2\varepsilon-\alpha}{\alpha r}\right)\frac{df_{n}(r)}{dr}
+\left(\frac{2\varepsilon\beta+\alpha\beta-2\beta^{2}}{r\alpha^{2}}\right)f_{n}(r),
\end{equation}
where
\begin{equation}
-\tau(r)=2\beta r-2\varepsilon-\alpha, \quad \sigma=\alpha r, \quad
-\gamma_n=\left(\frac{2\varepsilon\beta+\alpha\beta-2\beta^{2}}{\alpha}\right).
\end{equation}
Using Eq. (15), the eigenvalues turn out to be
\begin{equation}\label{energym}
\varepsilon_{n}=\beta-(n+\frac{1}{2})\alpha
\hspace{1cm} n=0,1,2,3,\cdots.
\end{equation}

For the deformed H\`{u}lten potential case, if we take the
transformed equation, Eq.\,(18) of Ref.\,[16] can be rewritten as

\begin{equation}\label{aimschr}
\frac{d^{2}f(r)}{dr^{2}}=\left[\frac{(2\varepsilon+3)q r-(2\varepsilon+1)}{r(1-qr)}\right]
\frac{df(r)}{dr}+\left[\frac{2\varepsilon q+q-\beta^{2}}{r(1-qr)} \right]f(r).
\end{equation}
If we compare this equation with Eq. (6), we obtain the functions
\begin{equation}
-\tau(r)=(2\varepsilon+3)q r-(2\varepsilon+1),  \quad
\sigma(r)=-r(1-qr), \quad  -\gamma_n=2\varepsilon q+q-\beta^{2}.
\end{equation}
Using Eq. (15), the eigenvalues turn out to be
\begin{equation}\label{energy}
\varepsilon_{n}=-\frac{1}{(2n+2)}\left(\frac{q(n+1)^{2}-\beta^{2}}{q}\right),\hspace{1cm} n=0,1,2,3,\cdots.
\end{equation}

In the Kratzer potential case, if we take the transformed equation,
Eq. (14) of Ref.\,[17] can be rewritten as

\begin{equation}\label{diffaim}
  f^{''}(r)=2\left(\varepsilon-\frac{(\Lambda+1)}{r}\right)f^{'}(r)+\left( \frac{2(\Lambda+1)\varepsilon-A}{r}\right)f(r).
\end{equation}
If we compare this equation with Eq. (6), we obtain the functions
\begin{equation}
-\tau(r)=2\left(\varepsilon r-(\Lambda+1)\right),  \quad \sigma=r,
\quad  -\gamma_n=2(\Lambda+1)\varepsilon-A.
\end{equation}

Using Eq. (15), the eigenvalues turn out to be

\begin{equation}\label{energy2}
\varepsilon_{n}=-\frac{1}{(2n+2)}\left(\frac{q(n+1)^{2}-\beta^{2}}{q}\right),\hspace{1cm} n=0,1,2,3,\cdots
\end{equation}

Next, we discuss the eigenfunctions. If we insert Eq. (15) into Eq.
(7) and use Eqs. (2) and (3), we can obtain the wave functions as
follows:
\begin{eqnarray}\label{waveFst}
y_0(r) &=& 1, \\
y_1(r) &=& \tau,\\
y_2(r) &=& \tau^2+\tau\sigma'+\tau'\sigma+\sigma\sigma'',\\
y_3(r) &=& \tau^3+3\tau^2\sigma'+2\tau\sigma'^2+3\tau\tau'\sigma+4\tau'\sigma\sigma'+5\tau\sigma\sigma''+6\sigma\sigma'\sigma'', \\
\cdots  . \nonumber
\end{eqnarray}
These iterative functions can be generalized in a closed form as follows:

\begin{equation}\label{waveRodrigues}
y_n(r)=\frac{1}{\rho(r)}\frac{d^n}{ds^n}\left[\sigma(r)^n \rho(r)\right],\hspace{1cm} n=0,1,2,3,\cdots,
\end{equation}

where $\rho(r)$ has to satisfy$^{[3]}$
\begin{equation}\label{rhos}
\left[\sigma(r)\rho(r)\right]'=\tau(r)\rho(r) .
\end{equation}

If we apply Eq. (36) to the deformed Hulten potential case as an
example, we obtain

\begin{eqnarray}\label{waveFst1}
y_0(r) &=& 1, \\
y_1(r) &=&(2\varepsilon_1+3)q r-(2\varepsilon_1+1),\\
y_2(r) &=& 2 \left[(2 \varepsilon_2^2+3 \varepsilon_2)+(\varepsilon_2+2) (2 \varepsilon_2+5) q^2 r^2-4
   (\varepsilon_2+1) (\varepsilon_2+2) q r+1\right],\\
y_3(r) &=& 2(\varepsilon_3+3) (2 \varepsilon_3 +5) (2 \varepsilon_3 +7)q^3 r^3-6(\varepsilon_3+3) (2 \varepsilon_3 +3) (2 \varepsilon_3 +5) q^2 r^2+\\
   &&+2(3 \varepsilon_3 (4 \varepsilon_3(\varepsilon_3 +5)+31)+45)q r-2 (\varepsilon_3  (4 \varepsilon_3  (\varepsilon_3+3)+11)+3), \nonumber\\
\cdots . \nonumber
\end{eqnarray}
These equations can be written in a generalized way as a hypergeometric function,

\begin{equation}\label{waveHyper}
y_n(r)=(-1)^n\frac{\Gamma (2\varepsilon_n+n+1) \,}{\Gamma (2 \varepsilon_n+1)}) _2F_1\left(-n,\left(2 \varepsilon_n+n+2\right);\{2 \varepsilon_n+1\};q r\right) .
\end{equation}

If Eq. (42) is compared with the results of Ref.\,[16], it can be
seen that they are exactly the same eigenfunctions.

In summary, we have derived a formula for the analytically solvable
potentials, which is given in Eq. (15). This formula replaces the
iterative calculations and reduces the calculation workload in the
original AIM formulation to find the energy eigenvalues. We have
also shown that there is a close connection between the AIM and the
Nikiforov--Uvarov methods to solve differential equations.

%
\begin{table}
\begin{tabular}{lcccc} \hline
Differential equation &  $-\tau$&$\sigma$&$-\gamma_n$&Eigenvalue\\
\hline
Cauchy-Euler$^1$& ${\alpha(r-b)}$ & ${(r-a)^2}$& $\beta$ &  $\beta=n(n-1-\alpha)$ \\ 
Hermite$^{2a}$  & $2r$ & $1$& $-2k$ &  $k=n$ \\ 
Hermite$^{2b}$& $ar+b$ & $1$& $c$ &  $c=- na$ \\ 
Laguerre  & $r-1$ & $r$& $a$ &  $a=-n$ \\ 
Confluent$^3$  & $br-c$ & ${r}$& $a$ &  $a=-nb$ \\ 
Hypergeometric  & $(a+b+1)r-c$ & ${r(1-r)}$& $ ab $ &  $a=-n~ or~ b=-n$ \\ 
Legendre  & ${-2r}$ & ${r^2-1}$& ${m(m+1)}$ &  $m=n$ \\ 
Jacobi  & ${(\alpha+\beta+2)r+\beta+\alpha}$ & ${1-r^2}$& $-\gamma$ &  $\gamma=n(n+\alpha+\beta+1)$ \\ 
Chebyshev$^{4a}$& ${r}$ & ${1-r^2}$& $-m$ &  $m=n^2$ \\ 
Chebyshev$^{4b}$& ${3r}$ & ${1-r^2}$& $-m$ &  $m=n(n+2)$ \\ 
Gegenbauer & ${(1+2k)r}$ & ${(1-r^2)}$& $-\lambda$ &  $\lambda=n(n+2k)$ \\ 
Hyperspherical & ${2(1+k)r}$ & ${(1-r^2)}$& $-\lambda$ &  $\lambda=n(n+1+2k)$ \\ 
\textmd{Bessel$^{5a}$} & ${-2(r+1)}$ & ${r^2}$& $\gamma$ &  $\gamma=n(n+1)$ \\ 
Generalized Bessel$^{5b}$& ${-(ar+b)}$ & ${r^2}$& $\gamma$ &
$\gamma=n(n+a-1)$ \\\hline
\end{tabular}
\caption{Application of Eq. (15) to classical differential equations
which give polynomial solutions.} \label{table1}
\end{table}
\end{document}